%% file: main.tex
\documentclass[runningheads,orivec]{llncs}
\usepackage[T1]{fontenc}
\usepackage{amsmath}
\usepackage{amssymb}

\usepackage{graphicx,verbatim}
\usepackage{hyperref}
\hypersetup{
    colorlinks = true,
    allcolors  = blue
}
\usepackage{color}

\usepackage{bookmark}
\usepackage{booktabs}
\usepackage{csquotes}
\usepackage{cleveref}
\usepackage{subcaption}
\usepackage{siunitx}
\usepackage{threeparttable}

\usepackage[toc,acronym,nomain,nopostdot,nonumberlist]{glossaries}
\setacronymstyle{long-short}
\loadglsentries{glossary} 		% the glossary .tex file
\glsdisablehyper

\begin{document}

\title{End-to-end Cortical Surface Reconstruction from Clinical Magnetic Resonance Images} % of any Contrast and Resolution
\titlerunning{End-to-end Cortical Surface Reconstruction from Clinical MRI}

\begin{comment}  %% Removed for anonymized MICCAI 2025 submission
\author{First Author\inst{1}\orcidID{0000-1111-2222-3333} \and
Second Author\inst{2,3}\orcidID{1111-2222-3333-4444} \and
Third Author\inst{3}\orcidID{2222--3333-4444-5555}}
%
\authorrunning{J. D. Nielsen et al.}
\institute{Princeton University, Princeton NJ 08544, USA \and
Springer Heidelberg, Tiergartenstr. 17, 69121 Heidelberg, Germany
\email{lncs@springer.com}\\
\url{http://www.springer.com/gp/computer-science/lncs} \and
ABC Institute, Rupert-Karls-University Heidelberg, Heidelberg, Germany\\
\email{\{abc,lncs\}@uni-heidelberg.de}}

\end{comment}

\author{
Jesper Duemose Nielsen\inst{1} \and
Karthik Gopinath\inst{2} \and
Andrew Hoopes\inst{2,3} \and
Adrian Dalca\inst{2,3,4} \and
Colin Magdamo\inst{5} \and
Steven Arnold\inst{5} \and
Sudeshna Das\inst{2,5} \and
Axel Thielscher\inst{1,6} \and
Juan Eugenio Iglesias\inst{2,3,4,7} \and
Oula Puonti\inst{1,2}
}
\authorrunning{J. D. Nielsen et al.}

\institute{
Danish Research Centre for Magnetic Resonance, Department of Radiology and Nuclear Medicine, Copenhagen University Hospital -- Amager and Hvidovre, Copenhagen, Denmark \and
Athinoula A. Martinos Center for Biomedical Imaging, Massachusetts General Hospital, Charlestown, MA, USA \and
Computer Science and Artificial Intelligence Lab, Massachusetts Institute of Technology, Cambridge, MA, USA \and
Department of Radiology, Harvard Medical School, Boston, MA, USA \and
Department of Neurology, Massachusetts General Hospital, Charlestown, MA, USA \and
Department of Health Technology, Technical University of Denmark, Kgs. Lyngby, Denmark \and
Hawkes Institute, University College London, London, UK
}

\maketitle              % typeset the header of the contribution

\begin{abstract} % 150-250 words

% introduction
Surface-based cortical analysis is valuable for a variety of neuroimaging tasks, such as spatial normalization, parcellation, and \gls{gm} thickness estimation. However, most tools for estimating cortical surfaces work exclusively on scans with at least 1 mm isotropic resolution and are tuned to a specific \gls{mr} contrast, often \gls{t1w}. This precludes application using most clinical \gls{mr} scans, which are very heterogeneous in terms of contrast and resolution.
% method
Here, we use synthetic domain-randomized data to train the first neural network for explicit estimation of cortical surfaces from scans of any contrast and resolution, without retraining.  Our method deforms a template mesh to the \gls{wm} surface, which guarantees topological correctness. This mesh is further deformed to estimate the \gls{gm} surface.
We compare our method to \gls{rac}, an implicit surface reconstruction method which is currently the only other tool capable of processing heterogeneous clinical \gls{mr} scans, on  ADNI and a large clinical dataset (n=1,332). We show a $\sim 50$ \% reduction in cortical thickness error (from 0.50 to 0.24 mm) with respect to \gls{rac} and better recovery of the aging-related cortical thinning patterns detected by FreeSurfer on high-resolution \gls{t1w} scans.
Our method enables fast and accurate surface reconstruction of clinical scans, allowing studies (1)~with sample sizes far beyond what is feasible in a research setting, and (2)~of clinical populations that are difficult to enroll in research studies.
%Additionally, it is faster and more accurate than previous methods. 
The code is publicly available at \url{https://github.com/simnibs/brainnet}.

% Time
%In order to perform downstream tasks, the surfaces still need to be spherically registered to a template. This step takes approximately 30 minutes of which the majority of the time is used for the spherical mapping which can be circumvented by exploiting the fact that we are deforming a template surface which has a predefined mapping to the sphere. Hence, only the spherical registration step is required.

%\gls{rac} whose runtime is approximately three hours. Most of this time is spent on extracting and fixing the surfaces. Here, we are able to generate cortical surfaces for both hemispheres in less than a second using a GPU or approximately one/two/three minutes on a CPU.

%The network has learned to deform a topologically correct template surface so as to fit the input image. Hence, the surfaces are guaranteed to be topologically correct.

% The neural network consists of two parts. The first part is a convolutional neural network which operates on the input image to estimate feature maps from which a convolutional graph neural network estimates a series of deformations that are applied to a set of template vertices. 

% Authors must provide keywords and are not allowed to remove this Keyword section.

\end{abstract}

\glsresetall

\section{Introduction}

% fmri:
% surface-based registrations
%   anticevic2009 : Comparing Surface-Based and Volume-Based Analyses of Functional Neuroimaging Data in Patients with Schizophrenia
% - resolution and statistical power
%   Oosterhof2011 : A comparison of volume-based and surface-based multi-voxel pattern analysis
%   Brodoehl2020 : Surface-based analysis increases the specificity of cortical activation patterns and connectivity results

% an argument for surface-based modeling over volume-based

% structural features of cerebral cortex are related to/reflect/... clinical conditions, aging, etc.

% surface-based analyses are better than volume-based at capturing
% - cortical thickness
% - increased sensitivity?

\paragraph{Background.}
The human cortex is a tightly folded sheet of neural tissue organized in distinct layers with characteristic cellular and microstructural features in different cortical areas. Morphometric analysis of cortical surfaces, namely the \gls{wm} and \gls{gm} (pial) surfaces, has improved our understanding of structural changes related to brain development~\cite{natu2021}, neurological and psychiatric disorders~\cite{kuperberg2003,querbes2009}, and aging~\cite{shaw2016}. For example, cortical thickness has proven especially powerful as a biomarker for detecting and tracking the development of dementia~\cite{querbes2009}. By limiting blurring across gyri and sulci, surface-based analysis also improves the sensitivity of functional \gls{mri} studies, possibly at the level of the individual layers~\cite{polimeni2018}.

% CLASSICAL APPROACHES
\paragraph{Related work.}
Classical surface reconstruction approaches~\cite{fischl2012,gaser2024,goebel2012,macdonald2000,shattuck2002,vanessen2001}
%BrainSuite~\cite{shattuck2002}, Caret~\cite{vanessen2001}, BrainVoyager~\cite{goebel2012} and Civet~\cite{macdonald2000} 
rely on iterative optimization and geometry processing algorithms.
%with carefully tuned hyperparameters. These methods generally work well but have long run times due to their iterative nature and their need for topology correction.
% FreeSurfer's surface pipeline
A representative example is FreeSurfer's widely used \texttt{recon-all} pipeline, where surface reconstruction starts from a volumetric segmentation of the \gls{wm} compartment, the boundary of which is first tessellated. Due in part to the finite resolution of the input image, the generated surface mesh may possess a number of implausible characteristics, e.g., bridges or holes, which are corrected \emph{post hoc} to ensure that the surface is topologically correct and homotopic (continuously deformable) to a sphere~\cite{fischl2001}. The corrected \gls{wm} surface is then deformed to optimally align with the local intensity gradients in the original image. Subsequently, the \gls{gm} surface is estimated by iteratively moving the vertices of the \gls{wm} surface outward while monitoring and fixing self-intersections until it aligns with the image gradient~\cite{dale1999}.

% DEEP LEARNING APPROACHES
Classical methods have long run times due to their iterative nature and their need for topology correction. More recent \gls{dl} methods can speed up cortical reconstruction. Here we distinguish two main approaches: implicit and explicit methods. The former estimate a function that implicitly defines a surface, typically evaluated on the image grid. The function can represent a tissue segmentation, as in FastSurfer~\cite{henschel2020}, or a real-valued \gls{sdf} whose zero level-set defines the surface~\cite{gopinath2024,cruz2021}. Importantly, as these methods do not predict a surface directly, the surface needs to be constructed using an isosurface extraction method such as marching cubes~\cite{lorensen1987}. The resulting surface may suffer from topological defects and post-processing is typically necessary, for example, to ensure that it is homotopic to a sphere. Explicit methods~\cite{bongratz2024,hoopes2022,santacruz2022} deform a template mesh directly and are potentially more accurate. They typically consist of two parts: a \gls{cnn}, to extract volumetric features, and a \gls{gcn}, which estimates deformation vectors for the surface vertices by sampling the volumetric features. An attractive property of explicit methods is that the predicted surfaces are guaranteed to be topologically correct, provided that the original template is. Consequently, no topology correction is needed---but self-intersections may still be present.

\paragraph{Limitations of current methods.}
The methods listed above are tuned to specific acquisitions, typically isotropic \gls{t1w} scans, either by training data (neural networks) or careful tuning of hyperparamters (classical methods). 
%While they excel in their specific domain, they lack versatility when applied to diverse datasets. 
This is generally not a problem in research, where 1 mm isotropic \gls{t1w} scans are commonly acquired. However, it is a severe limitation when analyzing clinical \gls{mri} sessions, which generally do not include such 1 mm \gls{t1w} volumes, but rather a plethora of scans with heterogeneous contrast and resolution, often with larger slice spacings~\cite{billot2023}. The ability to effectively process such diverse data would open up opportunities for studies (1)~with sample sizes far exceeding those typically achievable in a research setting, and (2)~of certain clinical populations that may be difficult to enroll in research studies (e.g., rare diseases).

\paragraph{Contribution.} 
Here, we present the first deep learning method for explicit cortical surface reconstruction of brain \gls{mri} scans of any contrast and resolution. We build on a domain randomization approach with synthetic data that has been successfully applied to volumetric tasks~\cite{billot2023,hoffmann2022,iglesias2023} and which enables processing of heterogeneous clinical scans without retraining. Our proposed method is considerably faster and more accurate than \gls{rac}~\cite{gopinath2024}, the only existing cortical reconstruction method that can also cope with heterogeneous \gls{mri} data. \gls{rac} also builds on domain randomization but, being implicit, it is slow (as it requires iterative surface fitting and topology correction) and its accuracy is limited by the discrete \gls{sdf}.

\section{Methods}
%We wish to be able to estimate cortical surfaces (\gls{wm} and \gls{gm}) from brain scans. Specifically, given an input image and a template surface, which has been co-registered to said image, we seek a model that deforms the vertices of the template surface such as to fit the input image. To simplify our task, we make a couple of assumptions. First, the input image should be aligned such that the linear part of its affine transformation (from voxel to world space) is a $3 \times 3$ identity matrix (i.e., \gls{ras} orientation, \qty{1}{\milli\meter\cubed} voxels). Second, the template surface (which exists in \gls{mni} space) needs to be aligned with the input image.

\subsection{Problem Formulation}
\begin{figure}[t]
    \centering
    \includegraphics[alt={Training framework}, width=\textwidth]{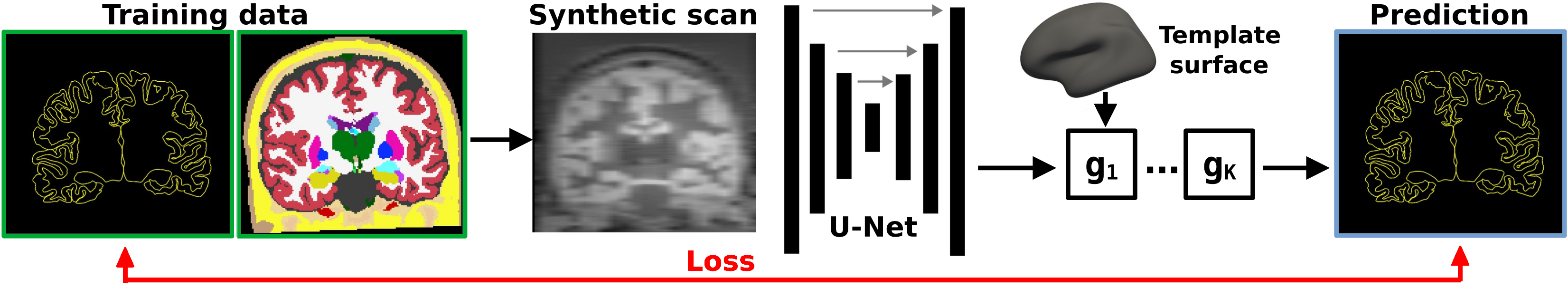}
    \caption{Overview of the synthetic data generation and training approach.}
    \label{fig:pipeline}
\end{figure}

Given an \gls{mri} scan, our goal is to estimate the \gls{wm} and \gls{gm} surfaces (see \cref{fig:pipeline} for an overview). These surfaces, which are 2D manifolds embedded in 3D Euclidean space, are defined by vertices and their connectivity. We use $M = \left\{ V, F \right\}$ to denote a surface mesh where $V \in \mathbb{R}^{\lvert V \rvert \times 3}$ and $F \in \mathbb{N}^{\lvert F \rvert \times 3}$ are the vertices and faces of the manifold, respectively. Here we assume that we are given the initial location of a set of vertices, which can be obtained from an affine registration of a template to the target scan, and we need to estimate, for each vertex, a (series of) displacement vector(s) that aligns these with the anatomical boundaries in the input scan. This problem can be described by an \gls{ode} of the form~\cite{santacruz2022}:
\begin{equation}
    \frac{d V(t)}{dt} = g\left(s\left[ f(I), V(t) \right]\right),
    \label{eq:surface-ode}
\end{equation}
where $I$ is the input scan, $t \in \left[ 0, 1 \right]$ is the time, and $V(t)$ are the vertex positions at time $t$. The function $f$ returns a number of features extracted from the input scan, $s$ is a function that samples $f(I)$ at $V(t)$, and $g$ uses these (sampled) features to estimate the positional change of each vertex. The boundary conditions of this equation, i.e., the values of $V(0)$ and $V(1)$, are given as the position of the template surface and the final vertex positions, respectively.
To solve \cref{eq:surface-ode}, we discretize the time interval in $K$ steps and apply a simple forward Euler integration scheme with step size $h=1/K$,
\begin{equation}
    V(k+1) = V(k) + h \times g(s\left[f(I), V(k)\right]).
    \label{eq:forward-euler}    
\end{equation}
%(It is also possible to apply more advanced Runge-Kutta methods at the expense of increased computational load and memory consumption.)

\subsection{Model Definition}
%In this work, we model the functions $f$ and $g$ using a neural network. Specifically, we use the TopoFit architecture presented in~\cite{hoopes2022} with minor modifications.

% f = UNet
Following~\cite{hoopes2022}, we model $f$ using a \gls{cnn} with a UNet architecture~\cite{ronneberger2015}. The UNet has five levels (encoder channels 16, 32, 64, 96, 128; decoder channels 96, 64, 64, 32) between which we do max-pooling with a kernel size of 2 in the encoder and upsampling with a factor of 2 in the decoder. At each level, $3 \times3 \times 3$ convolutions are applied followed by instance normalization and \gls{prelu} activation. The \gls{wm} surface is modeled by solving a recursive series of the function $g$, i.e.,
\begin{equation}
    V_{n+1} = g_{n+1}\left( s\left[ f(I), u\left( g_{n}\left( s\left[ f(I), V(t) \right]\right)\right)\right]\right),
\end{equation}
where $u$ is a mesh upsampling operation. That is, we sample the image features, $f(I)$ at $V(t)$, solve \cref{eq:forward-euler} $K$ times, upsample the surface, and repeat until the desired mesh resolution has been achieved.
Each $g$ is modeled with a \gls{gcn} with a UNet-like architecture, which has a maximum depth of four levels (encoder channels 64, 64, 64, 64, max-pooling; decoder channels 64, 64, 64, max-unpooling). It uses blocks of graph convolutions on the edges, instance normalization, and \gls{prelu} activation. Finally, a graph convolution layer is used to estimate a deformation vector. This UNet is repeated twice at each level, i.e., $K=2$. The initial template has 62 vertices ($n=0$) and we recurse until $n=6$ (245,762 vertices). Starting from the \gls{wm} surface, the \gls{gm} surface is placed using only a single $g$ since this displacement is relatively simple compared to predicting the \gls{wm} surface from the template positions. Here, $g$ is a simple linear layer with 32 channels followed by \gls{prelu} activation and another linear layer to predict the displacement vector. We repeat this process $K=10$ times.

\subsection{Training the Model}
%The following sections describe the datasets used during training, how they are processed, and what losses are minimized in order to estimate the model parameters.

\subsubsection{Datasets}
We train our model on multiple datasets (5898 total subjects; ages 7--97)~\cite{dimartino2014,milham2012,jackjr.2008,rowe2010,vogt2023,mayer2013,vanessen2013,carass2017,gollub2013,lamontagne2019} all of which include high-resolution (near 1 mm isotropic) \gls{t1w} scans. All scans were resampled to $1 \times 1 \times 1 \; \text{mm}^3$ and processed with FreeSurfer 7.4.1 to obtain the \gls{wm} and \gls{gm} surfaces and a brain segmentation. For generating synthetic \gls{mri} scans (see below), we simplify the volume segmentation by merging structures with similar intensity profiles (e.g., amygdala, cortex, hippocampus), and model the non-brain tissues using $k$-means clustering. Importantly, to account for the \gls{pv} effect close to the cortex, we encode the \gls{wm}, \gls{gm}, and \gls{csf} labels using the signed distance of a voxel to the nearest surface (\gls{wm} or \gls{gm}) (details below). The cortical surfaces, which are the training targets, were resampled to the template surface, which the model learns to deform, to allow calculation of distances between matched vertices (see \cref{sec:learning-parameters}). The template is co-registered to each subject using the Talairach registration from FreeSurfer. The dataset is split to train, test, and validation with fractions 0.8, 0.1, and 0.1.

\subsubsection{Domain Randomization}
\label{sec:data-synthesis}
Similar to~\cite{billot2023}, we generate synthetic \gls{mri} scans by sampling, for each voxel, intensity values from a Gaussian distribution conditioned on the label images. Specifically, we sample a mean and a standard deviation for each label, draw an image according to the mean, apply smoothing with a Gaussian kernel with random standard deviation to avoid sharp borders between regions with different labels, and finally add noise according to the standard deviation of each label. We force a minimum contrast between \gls{wm}, \gls{gm}, and \gls{csf} to stabilize the training. We model the \gls{pv} effect around the cortex by converting the signed distances to \gls{pv} fractions using a scaled sigmoid as the transfer function, $\text{PV}(d) = \frac{1}{1 + e^{-\rho d}}$, where $d$ is the distance and $\rho$ controls the steepness of the transfer function (high $\rho$, steeper \gls{pv}). Next, we apply a Gamma transform with a probability of 0.33 and a bias field with a probability of 0.75. Finally, we simulate a range of isotropic and anisotropic voxel resolutions along different dimensions. To handle different resolutions at test time,  we resample the input to $1 \times 1 \times 1 \; \text{mm}^3$ and min-max normalize the intensities.

\subsubsection{Learning}
\label{sec:learning-parameters}
During training we seek to find the network parameters which minimize a combination of data fidelity and regularization losses.
We use three fidelity terms that encourage accurate placement of the vertices. The first one is the symmetric chamfer distance,
\begin{equation}
    L_\text{chamfer}(P_x, P_y) = \dfrac{1}{\lvert P_x \rvert} \sum_{\mathbf{x} \in P_x} \min_{\mathbf{y} \in P_y} \lVert \mathbf{x}-\mathbf{y} \rVert ^ 2  + \dfrac{1}{\lvert P_y \rvert} \sum_{\mathbf{y} \in P_y} \min_{\mathbf{x} \in P_x} \lVert \mathbf{y}-\mathbf{x} \rVert ^ 2,
    \label{eq:chamfer}
\end{equation}
where $\mathbf{x}$ and $\mathbf{y}$ are points on the predicted and target surfaces, respectively. $P_x$ ($P_y$) denotes a set consisting of 100,000 points sampled from the predicted (target) surface and $\lvert \cdot \rvert$ its cardinality. The second fidelity term is a matched vertex distance which is useful to stabilize training in the initial phase,
\begin{equation}
    L_\text{matched}(M_x, M_y) = \dfrac{1}{\lvert V \rvert}\sum_{\mathbf{x} \in V_x, \mathbf{y} \in V_y} \lVert \mathbf{x} - \mathbf{y}\rVert^2.
\end{equation}
%, although it has been successfully employed previously~\cite{rickmann2023}. 
The third fidelity term, $L_\text{curv}(H_x,H_y)$, tries to match the curvature between the predicted and ground-truth surfaces. Specifically, we estimate the mean curvature using the discrete mean curvature normal (Laplace-Beltrami) operator~\cite{meyer2003},
and calculate the symmetric curvature loss as in \cref{eq:chamfer}, where the inputs are $H_x$ and $H_y$, which denote sets of mean curvature sampled at positions $P_x$ and $P_y$. Since the resampled surfaces are not always smooth, we apply Taubin smoothing~\cite{taubin1995} to the \emph{target surface} before computing its curvature. Additionally, we clip this estimate to percentile ranges $\left[ .001, .999 \right]$ and $\left[ .01, .99 \right]$ for the \gls{wm} and \gls{gm}.

We also use two regularization terms. First, we use a spring loss which encourages smoothness by penalizing inconsistencies of neighboring face normals:
\begin{equation}
    L_\text{spring}(M_x) = \dfrac{1}{\lvert E(F_x) \rvert}\sum_{i,j \in E(F_x)} \lVert \mathbf{n}_i - \mathbf{n}_j \rVert ^ 2.
\end{equation}
Here $\mathbf{n}_i$ denotes the normal of the $i$th face and $E(F_x)$ is the set of faces that share an edge. Second, we use an edge loss which encourages equilateral triangles by penalizing the variance of the normalized edge lengths, $\bar{E}(V_x)$:
\begin{equation}
    L_\text{edge}(M_x) = \dfrac{1}{\lvert \bar{E}(V_x) \rvert}\sum_{e \in \bar{E}(V_x)} \left( e - 1 \right)^2.
\end{equation}
The normalization is done to avoid shrinking the mesh, which might happen if the mean squared edge lengths are penalized directly.

During training, we gradually change the weights of each loss as indicated in the following by $(W_\text{start} \rightarrow W_\text{end})$. In the initial training phase, we put more emphasis on $L_\text{matched}$ $(1\rightarrow0)$ and $L_\text{spring}$ $(100 \rightarrow 0)$ which we gradually change to emphasize $L_\text{chamfer}$ $(1 \rightarrow 1)$ and $L_\text{curv}$ $(40 \rightarrow 2.5)$.
The optimization is done on an Nvidia RTX 6000 GPU (48 GB) using the AdamW optimizer with an initial learning rate of $10^{-4}$ which we decrease to $5\times10^{-5}$. In all our experiments, we train until the chamfer loss on the validation set plateaus which was approximately 160,000 iterations (33 hours).

\subsection{Evaluation}
We evaluate the performance of our model against \gls{rac}~\cite{gopinath2024}, the only existing method able to extract cortical surfaces from brain scans of arbitrary contrast and resolution. We use two evaluation metrics: (1)~cortical thickness estimation against a reference calculated with FreeSurfer on 1 mm isotropic \gls{t1w} scans, and (2)~sensitivity to age-related thickness changes.
We use two datasets for the comparisons. First, 200 subjects from the \gls{adni} GO/2 dataset (ages 56--89) which contains paired $1 \times 1 \times 1 \; \text{mm}^3$ \gls{t1w} scans and $0.85 \times 0.85 \times 5 \; \text{mm}^3$ \gls{flair} scans. We use the \gls{flair} scans as input and the FreeSurfer output on the \gls{t1w} (co-registered to the \gls{flair}) as ground truth.
The second dataset is a large clinical dataset consisting of 1,332 subjects (ages 18--90) acquired at Massachusetts General Hospital, which contains scans of varying contrasts (\gls{t1w}, T2w, \gls{flair}, DWI, etc.) and resolutions (slice spacings from 1--8 mm). Crucially, the \gls{mri} sessions for these subjects included high-resolution \gls{t1w} scans, which were processed with FreeSurfer to generate the ground truth.

\begin{figure}[t]
    \centering
    % \begin{subfigure}[b]{0.32\textwidth}
    %     \includegraphics[width=\linewidth]{357702_RA.png}
    %     % \caption*{\texttt{recon-all} (\gls{t1w})}
    % \end{subfigure}
    % \begin{subfigure}[b]{0.32\textwidth}
    %     \includegraphics[width=\linewidth]{357702_RAC_arrows.png}
    %     % \caption*{\gls{rac}}
    % \end{subfigure}
    % \begin{subfigure}[b]{0.32\textwidth}
    %     \includegraphics[width=\linewidth]{357702_TOPOFIT.png}
    %     % \caption*{Our method}
    % \end{subfigure}
    % \begin{subfigure}[b]{0.32\textwidth}
    %     \includegraphics[width=\linewidth]{357702_RA_mesh.png}
    %     \caption*{\texttt{recon-all} (\gls{t1w})}
    % \end{subfigure}
    % \begin{subfigure}[b]{0.32\textwidth}
    %     \includegraphics[width=\linewidth]{357702_RAC_mesh.png}
    %     \caption*{\gls{rac}}
    % \end{subfigure}
    % \begin{subfigure}[b]{0.32\textwidth}
    %     \includegraphics[width=\linewidth]{357702_TOPOFIT_mesh.png}
    %     \caption*{Our method}
    % \end{subfigure}
    \includegraphics[alt={Surface reconstruction example}, width=\textwidth]{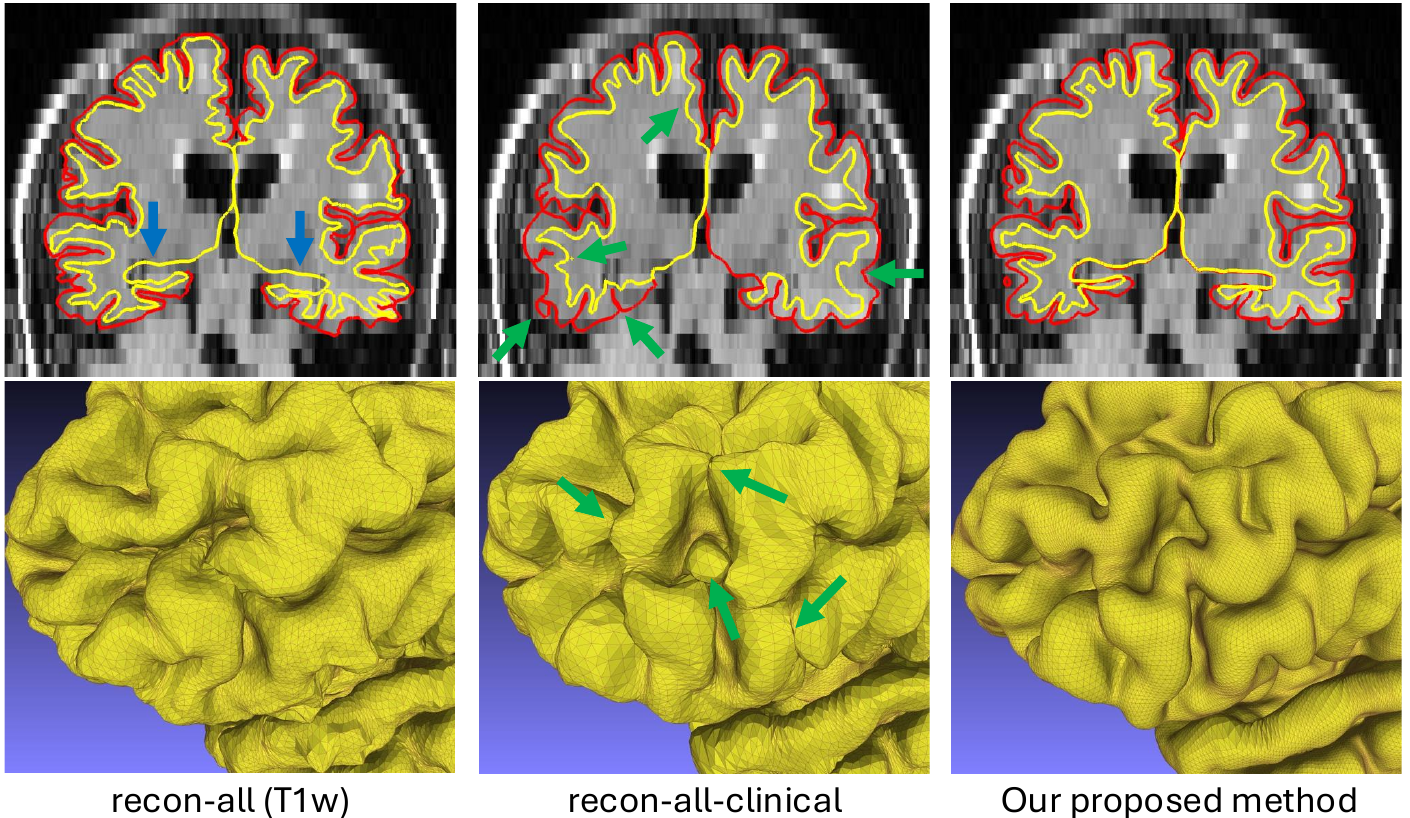}
    \caption{Cortical surface reconstruction on $0.85 \times 0.85 \times 5 \; \text{mm}^3$ \gls{flair} (\texttt{recon-all} reconstruction based on $1 \times 1 \times 1 \; \text{mm}^3$ \gls{t1w} for reference). Green arrows highlight regions of poor reconstruction by \gls{rac}. Note that \gls{rac} does not circumvent the hippocampus (blue arrows); this region, together with the medial wall, is masked out of the evaluation.}
    \label{fig:surface-example}
\end{figure}

\begin{figure}[t]
    \centering
    \includegraphics[alt={Absolute thickness errors for ADNI GO/2 and clinical dataset}, width=\textwidth]{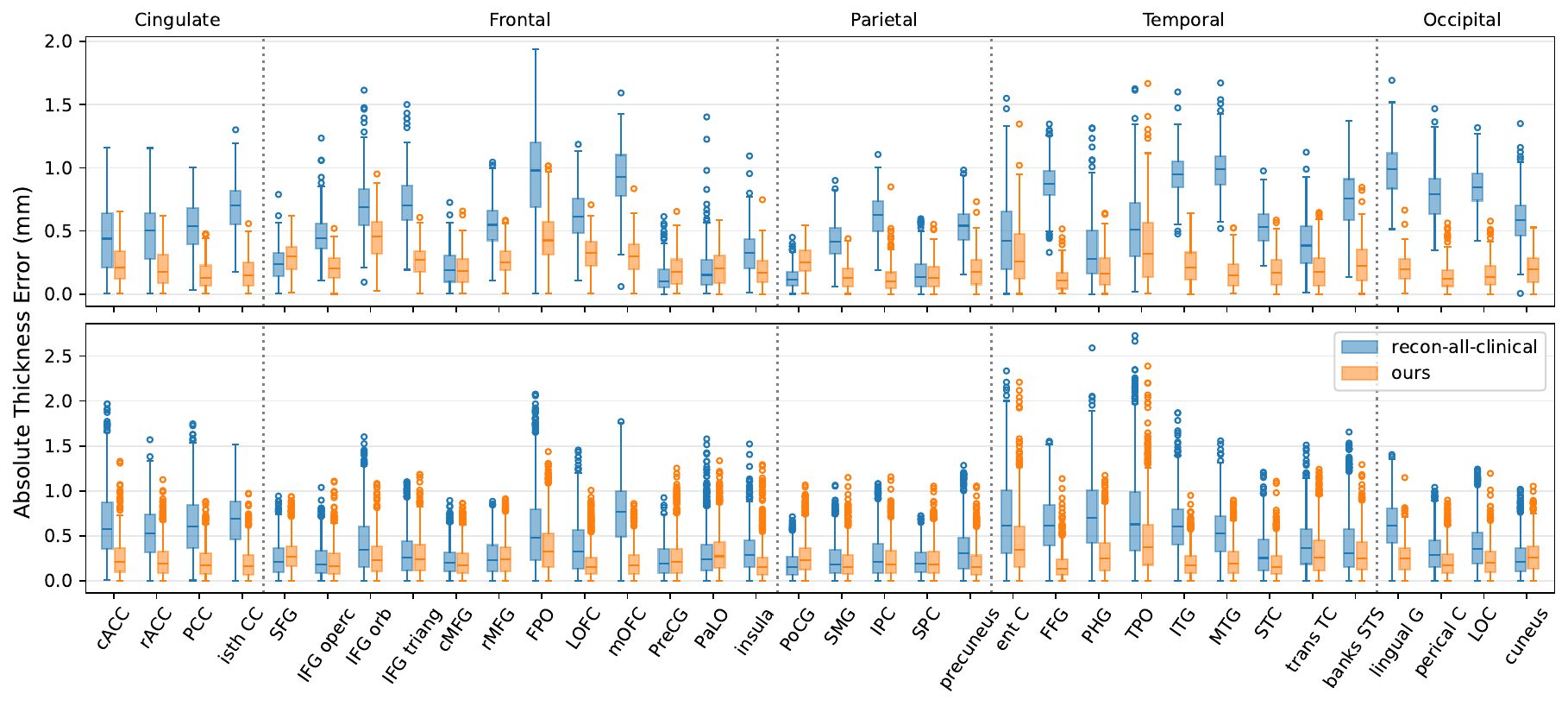}
    \caption{Box plots for absolute value of mean cortical thickness error on the axial \gls{flair} scans from \gls{adni} GO/2 (top) and the clinical dataset (bottom).}
    \label{fig:thickness}
\end{figure}
%          recon-all-clinical & 0.940 (0.079) & 0.926 (0.080) && 2.376 (0.207) & 2.689 (0.275) \\
%          ours & 0.800 (0.065) & 0.796 (0.071) && 1.942 (0.181) & 2.049 (0.189) \\
%         \bottomrule
\section{Results}
% In \cref{tab:acc} we report surface reconstruction errors from which it is clear that our method consistently outperforms \gls{rac}.
Our method outperforms \gls{rac} in terms of surface accuracy. The mean symmetric surface reconstruction errors for white and pial surfaces are 0.940 and 0.926 mm for \gls{rac} and 0.800 and 0.796 mm for our method. Likewise, the 90th percentile Hausdorff distances for white and pial surfaces are 2.376 and 2.689 mm for \gls{rac} and 1.942 and 2.049 mm for our method.

\Cref{fig:surface-example} shows a sample surface reconstruction from \gls{adni} GO/2 using the different approaches. Although the \gls{pv} effects on low-resolution data make accurate surface prediction an extremely challenging task, both \gls{rac} and the proposed approach successfully capture the general folding pattern. Our method does, however, recover smaller folds that are missed by \gls{rac} (green arrows).

\begin{figure}[h!]
    \centering
    \includegraphics[alt={Age-thickness relationship for ADNI GO/2}, width=\textwidth]{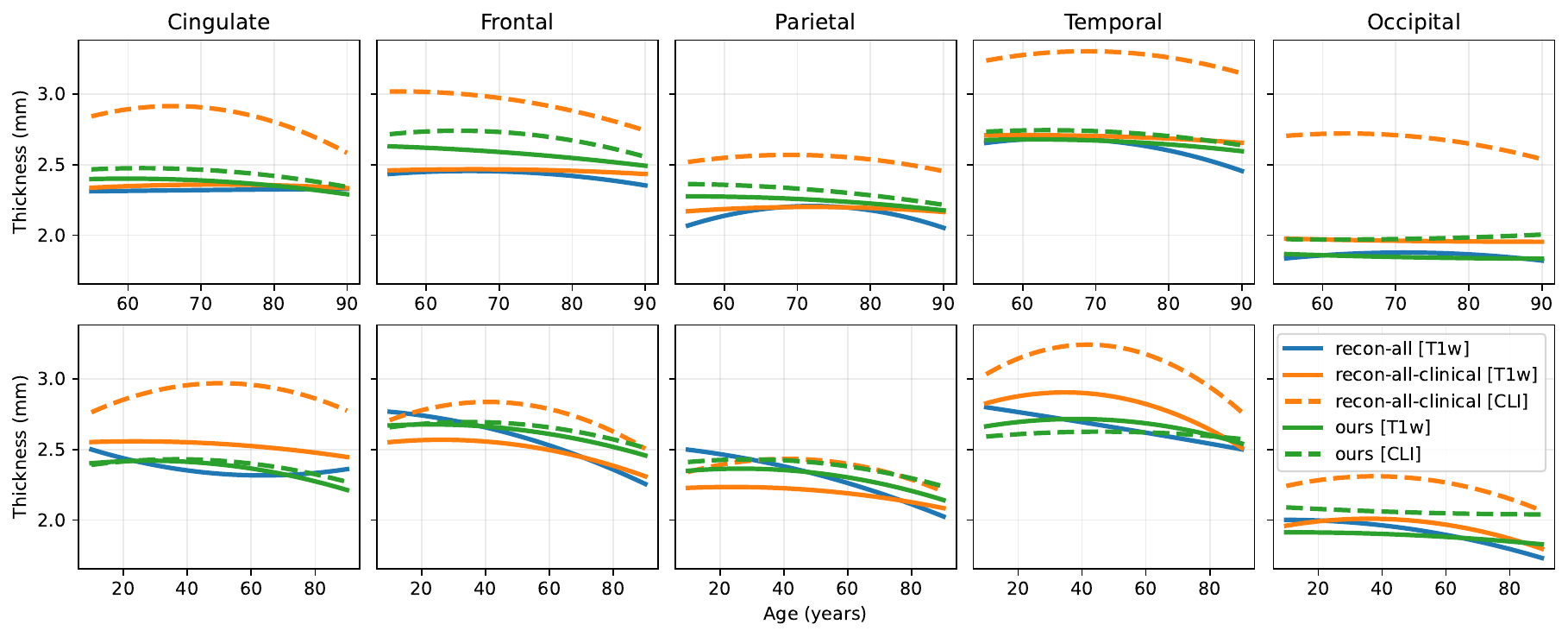}
    \caption{Cortical thickness variation with age on \gls{adni} GO/2 (top) and the clinical dataset (bottom). Solid (dashed) lines show the age trend using \gls{t1w} (clinical) scans estimated using a quadratic fit. Note the different ranges on the $x$ axis.}
    \label{fig:clinical-age-thickness}
\end{figure}

In \cref{fig:thickness}, we show the absolute value of mean thickness errors in different cortical areas. We see that our method produces more accurate thickness estimates than \gls{rac} in most regions. % -- all except the central gyri (Pre/PoCG). 
Remarkably, the average error is reduced by over 50 \% (from 0.50 to 0.24 mm).
Age-thickness relationships are shown in \cref{fig:clinical-age-thickness}. For both datasets, we find that the thickness estimates from our approach are overall less biased (offset on the $y$ axis) and that the trends are more similar to those from \texttt{recon-all} using high-resolution \gls{t1w} scans.

Finally, we note that the mean number of \gls{sif} is negligible for both approaches: \gls{rac} yields 0.001 \% in \gls{wm} and 0.002 \% in \gls{gm}, whereas our method yields 0.005 \% and 0.006 \%, respectively.

%The supplementary data shows a video of the surface reconstruction for an example low-resolution scan.
% The 95th percentile is \qty{0.003}{\percent} and \qty{0.006}{\percent} (\gls{rac}) and \qty{0.02}{\percent} and \qty{0.02}{\percent} (TopoFit).

%Run time for whole-brain prediction is approximately \qty{1}{\second} on a GPU and \qty{1}{\min} on a CPU.

% In \cref{fig:adni-surface-dist}, we show the surface accuracy
\iffalse
\begin{figure}
	\centering
	\begin{subfigure}[b]{0.39\textwidth}
		\includegraphics[height=5.5cm]{miccai_322922_topofit_flair_axial.png}
	\end{subfigure}
        \hfill
	\begin{subfigure}[b]{0.59\textwidth}
		\includegraphics[height=5.5cm]{miccai_322922_topofit_flair_coronal.png}
	\end{subfigure}
	\caption{Surfaces estimated on a 5 mm axial \gls{flair} scan from the \gls{adni} GO/2 dataset. \Gls{wm} (yellow), \gls{gm} (red).}
	\label{fig:surface-example}
\end{figure}
\fi

% \begin{table}
%     \centering
%     \begin{tabular}{
%     c@{\quad}
%     S[table-format = 1.3(1), separate-uncertainty]
%     S[table-format = 1.3(1), separate-uncertainty]
%     c
%     S[table-format = 1.3(1), separate-uncertainty]
%     S[table-format = 1.3(1), separate-uncertainty]
%     }
%         \toprule
%         & \multicolumn{2}{c}{ASSD} & & \multicolumn{2}{c}{HD90} \\
%          & {white} & {pial} & &{white} & {pial} \\
%          \cmidrule{2-3} \cmidrule{5-6}
%          recon-all-clinical & 0.940 (0.079) & 0.926 (0.080) && 2.376 (0.207) & 2.689 (0.275) \\
%          ours & 0.800 (0.065) & 0.796 (0.071) && 1.942 (0.181) & 2.049 (0.189) \\
%         \bottomrule
%     \end{tabular}
%     \caption{Surface reconstruction accuracy (in millimeter) on \gls{adni} GO/2 calculated using FreeSurfer surfaces as ground truth. ASSD (average symmetric surface distance), HD90 (90th percentile Hausdorff distance).}
%     \label{tab:acc}
% \end{table}

\section{Discussion}
We have presented a neural network, trained exclusively on synthetic data, capable of estimating cortical surfaces on scans with highly variable contrast and resolution. Compared to \gls{rac}, a state-of-the-art implicit approach, our method is more accurate, achieves a lower thickness estimation error, and can replicate aging effects on cortical thickness more reliably.
Our approach is also significantly faster. It runs in 1 s on a GPU and a few minutes on a CPU compared to 1--2 hours required by \gls{rac}.
The surfaces produced by our method also have a very low number of \gls{sif} even though no specific post-processing steps are performed.
Future work will explore the application of our method as a foundation model (i.e., potential for fine-tuning to other datasets), and seek to close the performance gap with approaches trained on 1 mm isotropic \gls{t1w} scans.
In combination with other \gls{dl}-based models (e.g., for spherical registration \cite{li2024}) this is a first step towards a fully learning-based surface pipeline that not only runs in a few seconds but can be applied to any kind of \gls{mri} scans, thus enabling retrospective studies of cortical morphometry at an unprecedented scale.

\begin{credits}
\subsubsection{\ackname} 
JDN and OP were supported by the Lundbeck Foundation (R360–2021–39).
AD was supported by the National Institutes of Health (NIH) (R01EB033773).
CM was supported by NIH (R01AG058063).
AT was supported by the Lundbeck Foundation (R313-2019-622), the German Research Foundation (DFG grants TH 1330/6-1 and TH 1330/7-1, part of Research Unit FOR 5429 ``MeMoSLAP''), NIH (1RF1MH117428-01A1) and European Union's Horizon Europe research and innovation programme (grant agreement No. 101071008).
JEI was supported by NIH -- BRAIN Initiative (1RF1MH123195, 1UM1MH130981), NIH -- National Institute of Aging (1R01AG070988, 1RF1AG080371, 1R21NS138995), and NIH -- NIBIB (1R01EB031114).

% \subsubsection{\discintname}
% It is now necessary to declare any competing interests or to specifically
% state that the authors have no competing interests. Please place the
% statement with a bold run-in heading in small font size beneath the
% (optional) acknowledgments\footnote{If EquinOCS, our proceedings submission
% system, is used, then the disclaimer can be provided directly in the system.},
% for example: The authors have no competing interests to declare that are
% relevant to the content of this article. Or: Author A has received research
% grants from Company W. Author B has received a speaker honorarium from
% Company X and owns stock in Company Y. Author C is a member of committee Z.
\end{credits}

\clearpage
\bibliographystyle{splncs04_mod}
\bibliography{references} % Entries are in the references.bib file

% @misc{fomin2020,
%   author = {V. Fomin and J. Anmol and S. Desroziers and J. Kriss and A. Tejani},
%   title = {High-level library to help with training neural networks in PyTorch},
%   year = {2020},
%   publisher = {GitHub},
%   journal = {GitHub repository},
%   howpublished = {\url{https://github.com/pytorch/ignite}},
% }

\end{document}

%% file: glossary.tex
% datasets
\newacronym{abide}{ABIDE}{Autism Brain Imaging Data Exchange}
\newacronym{aibl}{AIBL}{Australian Imaging Biomarkers and Lifestyle Study of Aging}
\newacronym{adni}{ADNI}{Alzheimer's Disease Neuroimaging Initiative}
\newacronym{chinese-hcp}{Chinese-HCP}{Chinese HCP}
\newacronym{cobre}{COBRE}{Center for Biomedical Research  Excellence}
\newacronym{hcp}{HCP}{Human Connectome Project}
\newacronym{isbi2015}{ISBI2015}{2015 International Symposium in Biomedical Imaging}
\newacronym{mcic}{MCIC}{MIND Clinical Imaging Consortium}
\newacronym{oasis3}{OASIS3}{Open Access Series of Imaging Studies (phase 3)}

\newacronym{adhd}{ADHD}{attention-deficit/hyperactivity disorder}

% models
\newacronym{gmm}{GMM}{Gaussian mixture model}
\newacronym{cnn}{CNN}{convolutional neural network}
\newacronym{gcn}{GCN}{graph convolutional neural network}
\newacronym{ode}{ODE}{ordinary differential equation}
\newacronym{prelu}{PReLU}{parametric rectified linear unit}
\newacronym{sdf}{SDF}{signed distance function}

% anatomy
\newacronym{ras}{RAS}{right-anterior-superior}

\newacronym{pv}{PV}{partial volume}

\newacronym{csf}{CSF}{cerebrospinal fluid}
\newacronym{gm}{GM}{gray matter}
\newacronym{wm}{WM}{white matter}

% imaging
\newacronym{ct}{CT}{computerized tomography}
\newacronym{mr}{MR}{magnetic resonance}
\newacronym{mri}{MRI}{magnetic resonance imaging}
\newacronym{t1w}{T1w}{T1-weighted}
\newacronym{flair}{FLAIR}{fluid-attenuated inversion recovery}
\newacronym{mni}{MNI}{Montreal Neurological Institute}

% tools
\newacronym{charm}{CHARM}{Complete Head Anatomy Reconstruction Method}
\newacronym{cat12}{CAT12}{Computational Anatomy Toolbox 12}
\newacronym{rac}{\texttt{RAC}}{recon-all-clinical}
% misc
%\newacronym{mgh}{MGH}{Massachusetts General Hospital} % no need to get the paper rejected for breaking anomymity!
\newacronym{sif}{SIF}{self-intersecting faces}
\newacronym{dl}{DL}{deep learning}